\def\year{2021}\relax
\documentclass[letterpaper]{article} 
\usepackage{aaai21}  
\usepackage{times}  
\usepackage{helvet} 
\usepackage{courier}  
\usepackage[hyphens]{url}  
\usepackage{graphicx} 
\usepackage{xcolor}

\usepackage{latexsym}
\usepackage{amsmath}
\usepackage{multirow}
\usepackage{multicol}
\usepackage{fancyhdr,graphicx,amssymb}
\usepackage{CJKutf8}
\usepackage{url}
\usepackage{booktabs}
\usepackage{natbib}  
\usepackage{caption} 
\title{CodeBLEU: a Method for Automatic Evaluation of Code Synthesis}
\begin{author}{
        Shuo Ren\textsuperscript{\rm 1},
        Daya Guo\textsuperscript{\rm 2},
        Shuai Lu\textsuperscript{\rm 3},
        Long Zhou\textsuperscript{\rm 4},
        Shujie Liu\textsuperscript{\rm 4},\\
        Duyu Tang\textsuperscript{\rm 4},
        Neel Sundaresan\textsuperscript{\rm4},
        Ming Zhou\textsuperscript{\rm 4},
        Ambrosio Blanco\textsuperscript{\rm 4},
        Shuai Ma\textsuperscript{\rm 1}\\
}
\end{author}
\affiliations{
    \textsuperscript{\rm 1}SKLSDE Lab, Beihang University; Beijing Advanced Innovation Center for Big Data and Brain Computing \\
    \textsuperscript{\rm 2}Sun Yat-sen University  \textsuperscript{\rm 3}Peking University \textsuperscript{\rm 4}Microsoft \\
    \textsuperscript{\rm 1}\{shuoren, mashuai\}@buaa.edu.cn \textsuperscript{\rm 2}guody5@mail2.sysu.edu.cn \textsuperscript{\rm 3}lushuai96@pku.edu.cn\\ \textsuperscript{\rm 4}\{Long.Zhou, shujliu, dutang, neels, mingzhou, ambrob\}@microsoft.com
}
\newcommand{\tl}[1]{\multicolumn{1}{l|}{#1}}
\newcommand{\tll}[1]{\multicolumn{1}{l}{#1}}

\begin{document}
\begin{CJK*}{UTF8}{gbsn}
\maketitle

\begin{abstract}
Evaluation metrics play a vital role in the growth of an area as it defines the standard of distinguishing between good and bad models. In the area of code synthesis, the commonly used evaluation metric is BLEU or perfect accuracy, but they are not suitable enough to evaluate codes, because BLEU is originally designed to evaluate natural language, neglecting important syntactic and semantic features of codes, and perfect accuracy is too strict thus it underestimates different outputs with the same semantic logic. To remedy this, we introduce a new automatic evaluation metric, dubbed CodeBLEU. It absorbs the strength of BLEU in the n-gram match, and further injects code syntax via abstract syntax trees (AST) and code semantics via data-flow.
We conduct experiments by evaluating the correlation coefficient between CodeBLEU and quality scores assigned by the programmers on three code synthesis tasks, i.e., text-to-code, code translation, and code refinement. Experimental results show that, our proposed CodeBLEU can achieve a better correlation with programmer assigned scores compared with BLEU and accuracy. 
\end{abstract}

\section{Introduction}
A suitable evaluation metric is important to push forward the research of an area, such as BLEU \cite{papineni2002bleu} and ROUGE \cite{lin-2004-rouge} for machine translation and text summarization. Along with the rapid progress of code synthesis such as text-to-code synthesis, code translation and code change prediction \cite{karaivanov2014phrase,oda2015learning,barone2017parallel,chen2018tree,kanade2019pre,husain2019codesearchnet,feng2020codebert,dinella2020hoppity,lachaux2020unsupervised}, different automatic evaluation methods for code synthesis are leveraged, including n-gram accuracy \cite{karaivanov2014phrase}, perfect accuracy \cite{chen2018tree}, and computational accuracy \cite{lachaux2020unsupervised}.
The n-gram accuracy (e.g. 4-gram BLEU) is the most popular evaluation method for code synthesis \cite{karaivanov2014phrase,barone2017parallel}, based on the token overlapping between the hypothesis and the reference. 
The perfect accuracy calculates the percentage of the predicted target programs that are exactly the same as the ground truth \cite{chen2018tree}.
The recently proposed computational accuracy \cite{lachaux2020unsupervised}, evaluates whether the hypothesis function generates the same outputs as the reference given the same inputs.

However, the above evaluation approaches still face many drawbacks.
First, the n-gram accuracy does not take into account the grammatical and logical correctness, resulting in favoring candidates with high n-gram accuracy and serious logical errors.
Second, the perfect accuracy is too strict, and underestimates different outputs with the same semantic logic.
Third, the computational accuracy is weak in universality and practicability, since it should be designed for different programming languages, as well as specific compilers and the desired computing resource.


In order to deal with that, in this paper, we propose a new evaluation metric CodeBLEU, considering information from not only the shallow (n-gram) match, but also the syntactic match and the semantic match. 
More specifically, the n-gram match assigns different weights for different n-grams, the syntactic match considers the abstract syntax tree (AST) information in the evaluation score by matching the sub-trees, and the semantic match uses data-flow structure to measure the semantic similarity.
CodeBLEU is a weighted combination of the original BLEU, the weighted n-gram match, the syntactic AST match, and the semantic data-flow match.

We conduct massive experiments to evaluate the effectiveness of CodeBLEU and the correlation coefficient between CodeBLEU scores and human evaluation scores in three code synthesis tasks including text-to-code synthesis, code translation, and code refinement. 
Experimental results demonstrate that CodeBLEU can significantly differentiate the systems’ performance and achieve better correlation with the quality scores given by programmers than the popularly used BLEU.
We hope that our proposed CodeBLEU can accelerate the R\&D cycle of code synthesis tasks. 

\section{Why not BLEU?}
 In this section we will briefly introduce BLEU, and analyze its merits and demerits when applying it to code synthesis. 


\subsection{BLEU for Machine Translation}

Machine translation, which uses computers to realize automatic translation between languages, is first proposed by Warren Weaver as early as 1949 \cite{weaver1955translation}. Since then, machine translation quality has not significantly improved until the automatic evaluation metric (BLEU) is proposed in 2002 \cite{papineni2002bleu}. 
The appearance of BLEU makes it possible to automatically train and optimize the machine translation systems and speeds up the research process of machine translation. 

BLEU measures how well a candidate translation matches a set of translation references by calculating the percentage of n-grams overlapped between them. Besides, the brevity penalty is introduced to punish the candidates with a very short length, so it is hard for the MT system to cheat the evaluation metric by finding a way to change the output that the BLEU score goes up, but the translation quality doesn't. 

\subsection{Code vs Natural Language}

Although the BLEU achieves great success in the evaluation of machine translation and greatly encourages the research in this area, BLEU is not suitable for the evaluation of code synthesis without considering the characteristics of the programming language. A natural language is any language that has evolved naturally in humans through use and repetition, but code is artificially designed to produce various kinds of output. There are three big differences between them.

(1) \textbf{Limited keywords vs. millions of words}. Different from natural languages with a huge vocabulary, code is designed by humans and uses a small number of keywords, i.e., the reserved words of programming languages.
Intuitively, keywords are more important than other words and the keywords match should gain a higher score.

(2) \textbf{Tree structure vs. sequential structure}. Humans usually speak and write from left to right, and the current mainstream models usually process natural languages as a sequence \cite{zhou2019sequence}, such as end-to-end neural machine translation \cite{sutskever2014sequence,bahdanau2014neural,vaswani2017attention}. In contrast, code has a natural tree structure and needs to be compiled according to their abstract syntax tree \cite{rabinovich2017abstract}. 
Therefore, how to evaluate the syntactic structure of code becomes particularly important.

(3) \textbf{Unique instructions vs. ambiguous semantic}. Word sense disambiguation is a basic research problem in natural language processing, because natural languages usually have ambiguous and variable semantic. However, code design is required to be unique, standardized and systematic, with unique and fixed instructions.
This feature makes it possible to evaluate the semantics of the code.

In summary, code is significantly different from natural languages, and BLEU is not suitable for code synthesis evaluation only considering the token match and ignoring the importance of keywords, syntactic accuracy, and semantic correctness. 
Therefore, we propose a new evaluation metric CodeBLEU, which will be introduced in the following.

\begin{figure*}[!h]
	\centering
	\includegraphics[width=14cm]{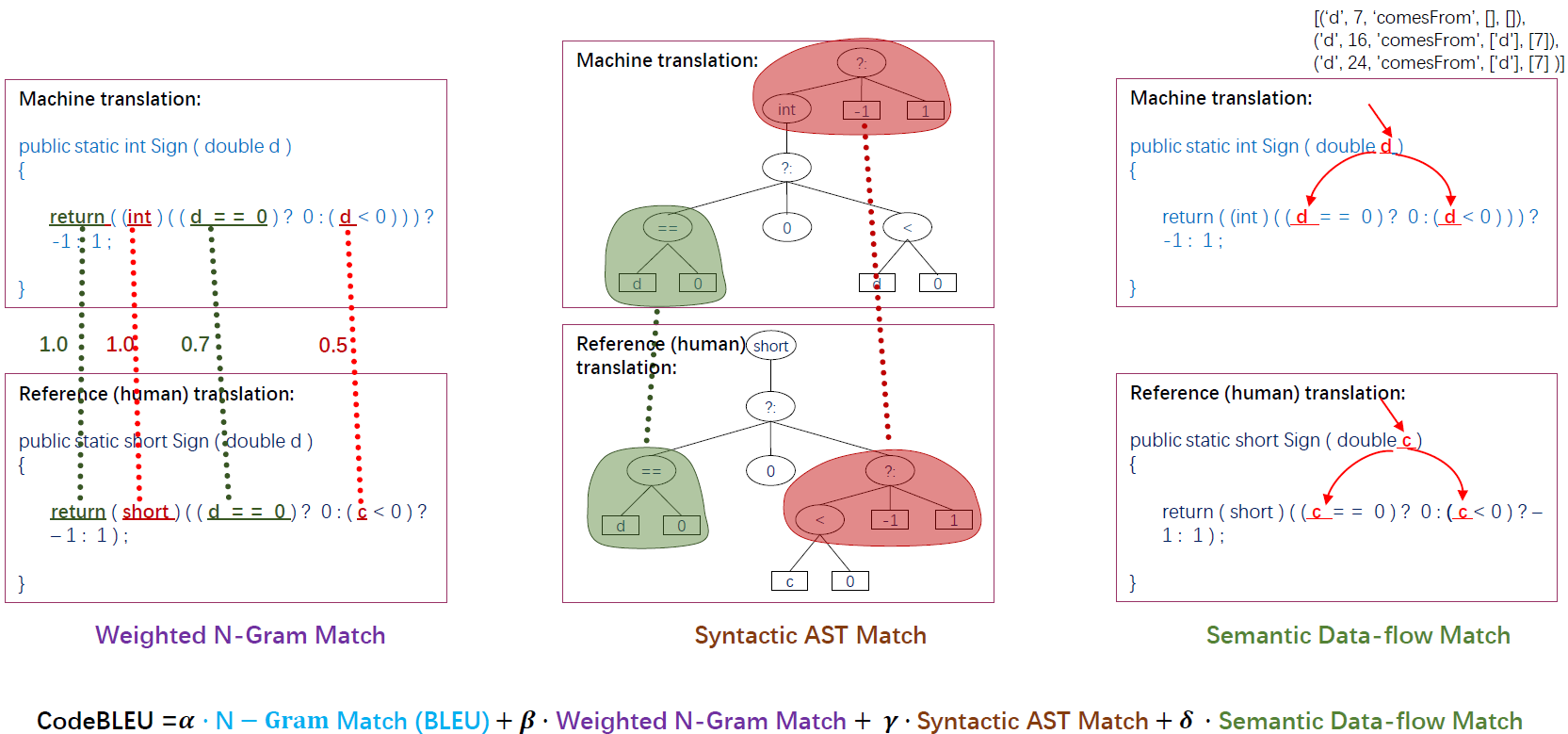}
	\caption{The proposed CodeBLEU, a weighted syntactic and semantic BLEU for code synthesis evaluation, consists of the original BLEU, the weighted n-gram match, the syntactic AST match, and the semantic data-flow match. 
	}\label{fig:score}
	\vskip -0.1in
\end{figure*}

\section{CodeBLEU}

In order to pay attention to the keywords, leverage the tree structure and consider the semantic logic information, we propose a new evaluation metric CodeBLEU defined as the weighted combination of four parts as shown in Figure \ref{fig:score}:
\begin{equation}
	\begin{aligned}
        \rm{CodeBLEU} =& \alpha \cdot \rm{BLEU} + \beta \cdot \rm{BLEU_{weight}} \\
        +& \gamma \cdot \rm{Match_{ast}} + \delta \cdot \rm{ Match_{df}}
	\end{aligned}
\label{eq:CodeBLEU}
\end{equation}
where $\rm{BLEU}$ is calculated by standard BLEU~\cite{papineni2002bleu}, 
$\rm{BLEU_{weight}}$ is the weighted n-gram match, obtained by comparing the hypothesis code and the reference code tokens with different weights (Sec. \ref{weight}), $\rm{Match_{ast}}$ is the syntactic AST match, exploring the syntactic information of code (Sec. \ref{syntax}), and $\rm{ Match_{df}}$ is the semantic data-flow match, considering the semantic similarity between the hypothesis and the reference (Sec. \ref{semantic}). The weighted n-gram match and the syntactic AST match are used to measure grammatical correctness, and the semantic data-flow match is used to calculate logic correctness.

\subsection{Weighted N-Gram Match}
\label{weight}

The original BLEU compares n-grams between the candidate and the reference, and calculates the ratio of matched n-grams.
Compared with natural languages which a huge vocabulary and a free word order, programming languages are manually designed and have only a few keywords such as ``int", ``public" and so on. 
Applying the traditional BLEU directly to code synthesis will ignore the importance of the keywords. 
Hence, we introduce the weighted n-gram match to assign different weights for different n-grams, so that the keywords may have higher weights, as shown in Figure \ref{fig:score}.

The weighted n-gram match precision is computed as:
\begin{equation}
	\begin{aligned}
		p_n = \frac{\sum\limits_{C\in {\rm{Candidates}}} \sum\limits_{i=1}^{l} \mu_n^i \cdot {\rm{Count_{clip}}} (C(i, i+n))}
		{\sum\limits_{C^{'}\in {\rm{Candidates}}} \sum\limits_{i=1}^{l} \mu_n^i \cdot {\rm{Count}} (C^{'}(i, i+n))}
	\end{aligned}
\end{equation}
where $n$ means the length of the n-gram, $C(i, i+n)$ is the n-gram from the position $i$ to the position $i+n$, and ${\rm{Count_{clip}}}(C(i, i+n))$ is the maximum number of n-grams co-occurring in a candidate code and a set of reference codes. $\mu_n^i$ denotes the weights of different keywords or n-gram.
In this paper, $\mu_n^i$ of the keywords is 5 times the weights of other tokens.
Next, following the brevity penalty of original BLEU, we also compute the brevity penalty BP:
$$ {\rm{BP}}=\left\{
\begin{aligned}
& 1           & if \ c > r \\
& e^{1-r/c}   & if \ c\leq r 
\end{aligned}
\right.
$$
where $c$ is the length of the candidate code
and $r$ is the effective reference corpus length. The weighted n-gram match score is calculated as:
\begin{equation}
	\begin{aligned}
        {\rm{BLEU_{weight}}} = {\rm{BP}} \cdot {\rm{exp}} (\sum\limits_{n=1}^{N} w_n {\rm{log}} p_n)
	\end{aligned}
\end{equation}
In our paper, the keywords are only considered in the uni-grams, so N and $w_n$ are equal to 1. Note that a keywords list is predefined for each programming language. 

\subsection{Syntactic AST Match}
\label{syntax}

In addition to the sequence-level matching, we also consider the syntactic information in CodeBLEU by matching the tree structure. 
Different from natural language, programming language has natural tree structures, such as the abstract syntax tree (AST). 
AST is a tree representation of the abstract syntactic structure of programming languages.
We can obtain all the sub-trees of the tree-sitter parsing result\footnote{https://github.com/tree-sitter/tree-sitter}, then calculate the accuracy by comparing the candidate and reference sub-trees.
In AST, each node denotes a construct occurring in the source code. The leaves of AST represent the names of the function and all the variables. However, we just want to use the syntactic structure of the codes, and the naming is not important, thus we leave out all the leave nodes in the original AST trees. 

As shown in the middle part of Figure 1, we extract all the sub-trees of the candidate and the reference ASTs respectively. Then we calculate the syntactic AST match score as:
\begin{equation}
	\begin{aligned}
        \rm{Match_{ast}} = Count_{clip}(T_{cand}) / Count(T_{ref})
	\end{aligned}
\end{equation}
where $\rm{Count(T_{ref})}$ is the total number of the reference subtrees, and $\rm{Count_{clip}(T_{cand})}$ is the number of the candidate subtrees that are matched the reference.
This score can evaluate code quality from a syntactic perspective, because grammatical errors such as token missing, data type errors can be captured by the difference between their ASTs.

\subsection{Semantic Data-flow Match}
\label{semantic}
\begin{figure}[!h]
	\centering
	\includegraphics[width=7.4cm]{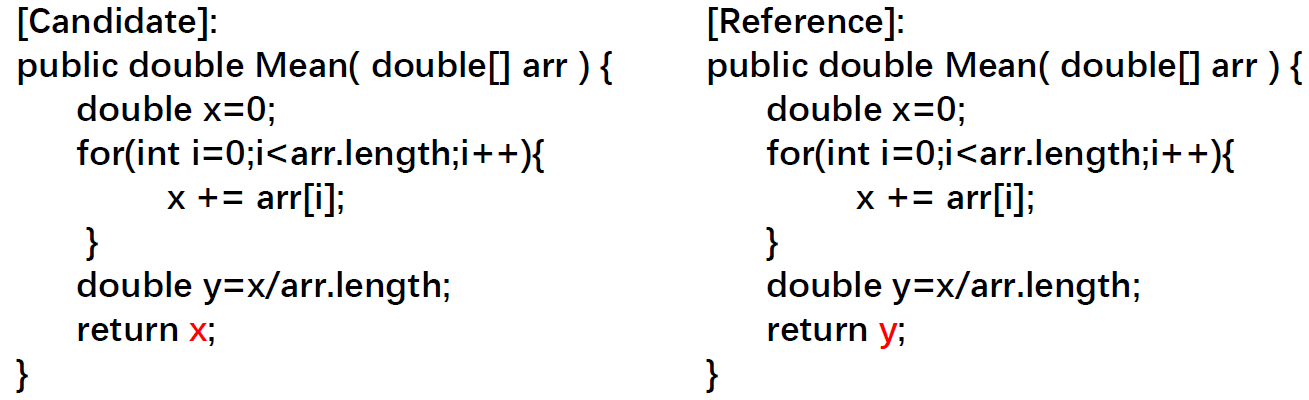}
	\caption{BLEU: 95.47; $\rm{ Match_{ast}}$: 100.
	}\label{fig:example_3}
	\vskip -0.12in
\end{figure}

In programming languages, the semantic of source code is highly relevant to the dependency relations among variables. Taking Figure \ref{fig:example_3} as an example, the function is to calculate the mean value of an array. Although the difference between the candidate and the reference is subtle ($return\ y\rightarrow return\ x$), their semantics are completely different. 
However, the weighted n-gram match and the syntactic AST match still give a high score since the two pieces of codes have the same AST and their tokens are highly overlapped.
Therefore, we also consider the semantic information in CodeBLEU. We use data-flow \cite{guo2020graphcodebert} to represent a source code as a graph, in which nodes represent variables and edges represent where the value of each variable comes from. Unlike AST, data-flows of the two codes are different in Figure \ref{fig:example_3} since their return values come from $x$ and $y$ respectively. Such a semantic graph can be used to measure the semantic match between the candidate and the reference. 

Based on the above, there are three steps to compute the semantic data-flow match score.

\textbf{Step 1}: Obtain the data-flow graphs for the candidate and the reference. Based on AST, we first utilize the leaves to identify variable sequence, denoted as $V = \{v_0, v_1, ..., v_m\}$. 
We then take each variable as a node of the graph and a directed edge $\epsilon = \langle{v_i,v_j}\rangle$ from $v_i$ to $v_j$ refers that the value of $j$-th variable comes from $i$-th variable.
The graph $\mathcal{G}(C) = (V;E)$ is used to represent relations among variables of the code $C$, as shown by the red arrows in Figure \ref{fig:score}.

\textbf{Step 2}: Normalize data-flow items. For simplicity and unity, we ignore the variable position and normalize their names. We collect all the variables in the data-flow items and rename them var\_$i$, where $i$ is the order of the variables appearing in all data-flow items. 

\textbf{Step 3}: Calculate the semantic data-flow match score as:
\begin{equation}
	\begin{aligned}
        \rm{Match_{df}} = Count_{clip}(DF_{cand}) / Count(DF_{ref})
	\end{aligned}
\end{equation}
where $\rm{Count(DF_{ref})}$ is the total number of the reference data-flows, and $\rm{Count_{clip}(DF_{cand})}$ is the number of matched candidate data-flows.

\begin{figure}[!th]
	\centering
	\includegraphics[width=8cm]{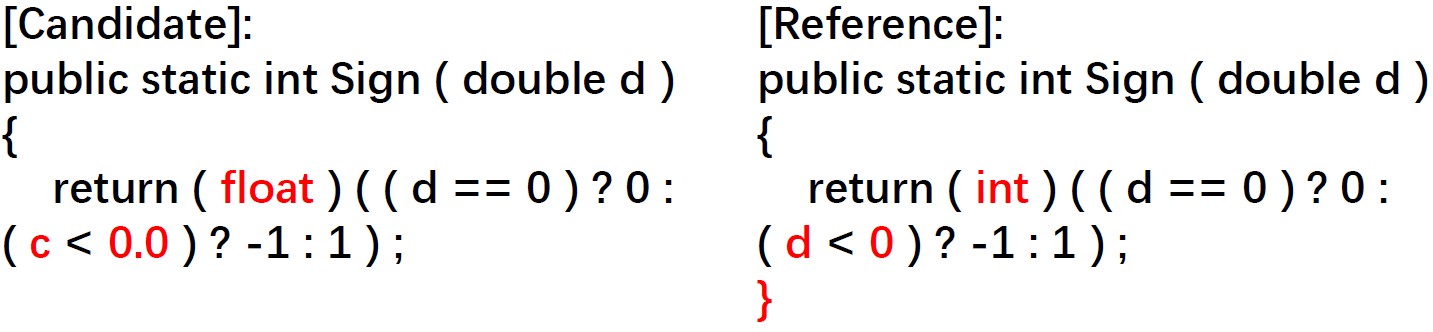}
	\caption{Example 1. BLEU: 75.43; CodeBLEU: 69.73.
	}\label{fig:example_1}
	\vskip -0.1in
\end{figure}

\subsection{Two Examples}
Here we will give two toy examples to show how to calculate CodeBLEU. Meanwhile, we show the qualitative advantages of CodeBLEU compared with the traditional BLEU score.

\subsubsection{Example 1}
The output candidate of a code synthesis system and the according reference are shown in Figure \ref{fig:example_1}.

In this example, there are four differences between the candidate and the reference, which are stressed with the red color. They are (1) the conversion type of the return value (``float" vs. ``int"); (2) the variable naming (``c" vs. ``d"); (3) the type of a constant (``0.0" and ``0"); (4) the missing token (``\}") in the candidate. This toy example is designed based on the background that the data type, the variable naming and the token missing tend to cause problems in reality. 

The CodeBLEU is calculated as follows: (1) First, we calculate the n-gram match score (BLEU, which is 75.43) given the candidate and the reference. (2) Then, we calculate the weighted n-gram match score for it. The weight assigned to the keywords "public, static, int, return, double" in the reference are 4 times more than that of the rest tokens. The resulting score is 74.91, lower than the BLEU score, penalizing the keyword error (``float" vs. ``int"). (3) The number of all sub-trees of the reference AST generated by tree-sitter is 21 and the hit number for the candidate is 13, so the syntactic AST match score is $13/21*100=61.90(\%)$. The data type errors in the candidate are penalized by the AST mismatch. (4) Three data-flows can be extracted from the reference AST, which are ``[(`var\_0', `comesFrom', []), (`var\_0', `comesFrom', [`var\_0'])], (`var\_0', `comesFrom', [`var\_0'])]", corresponding to the three variables ``d" in the reference. The first ``d" comes from no parent because it is in the parameter list. The second and the third ``d" come from the first ``d". The variable names are normalized and their positions are ignored according to Section \ref{semantic}. However, we can only extract two data-flows from the candidate AST , i.e., "[(`var\_0', `comesFrom', []), (`var\_0', `comesFrom', [`var\_0'])]" corresponding to the two ``d"s in this code. The variable ``c" is used before declaration so no data-flow is extracted for it. Therefore the data-flow match score is $2/3*100=66.67(\%)$. With $\alpha, \beta, \gamma, \delta = 0.25, 0.25, 0.25, 0.25$, the final CodeBLEU score is 69.73, which is lower than BLEU because CodeBLEU penalizes the keyword and semantic errors for the programming languages. 

\subsubsection{Example 2}

\begin{figure}[!h]
	\centering
	\includegraphics[width=8cm]{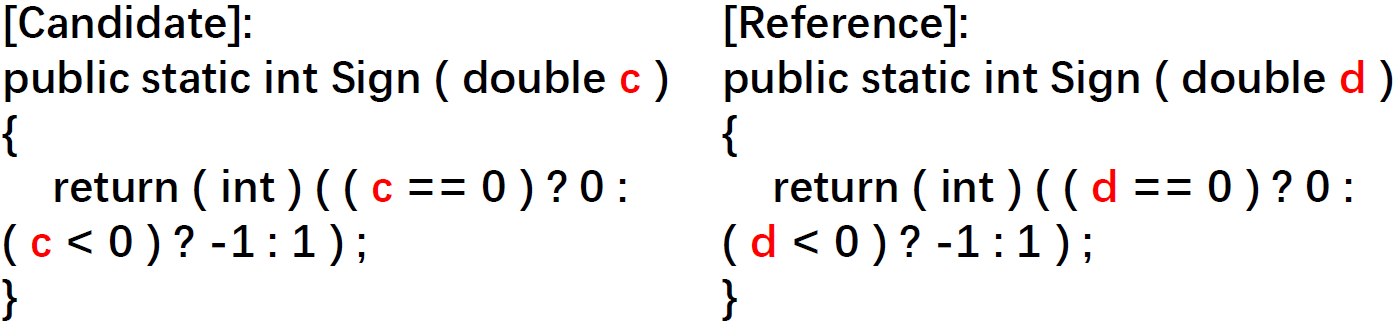}
	\caption{Example 2. BLEU: 68.14; CodeBLEU: 83.97.
	}\label{fig:example_2}
	\vskip -0.1in
\end{figure}

As shown in Figure \ref{fig:example_2}, in this example, there is no difference between the candidate and the reference except for the names of the local variables (``c" vs. ``d"). In the real scenario, the candidate is correct without doubt, and a human expert would give a score of 100. However, its BLEU score is only 75.71, which underestimates the quality of the candidate. With CodeBLEU, we have the weight n-gram match score of 76.46, the syntactic AST match score of 100 and the semantic data-flow match score of 100, the final CodeBLEU score being 88.04, which makes up for the underestimation of BLEU. 

From the two examples, we find that in some typical scenarios, CodeBLEU gives more reasonable scores than BLEU to evaluate the code synthesis output. In the experiment section, we will give the quantitative analysis, further showing the effectiveness of CodeBLEU.


\section{Experiments}
We conduct experiments on three code synthesis tasks, i.e., text-to-code (Java), code translation (from Java to {C\#}) and code refinement (Java). 
Previous work of these tasks uses BLEU or perfect accuracy (exactly match) for evaluation. In this paper, we will take the proposed CodeBLEU as the evaluation metric to see if CodeBLEU is more reasonable. For each task, we calculate the Pearson correlation coefficient to check the correlation between the scores given by our proposed CodeBLEU and the scores assigned by programmers (human evaluation scores). In the following subsections, we will first introduce the three tasks we used. Then we will give details of our experiment settings. Next, the experimental results will be shown and discussed. Finally, we will do an ablation study and investigate the influence of different components of CodeBLEU to the final results. 

\begin{table*}[ht]
\begin{center}
\setlength{\tabcolsep}{1mm}{
\begin{tabular}{c|c|c|c}
\toprule
Task & Text-to-code & Code translation & Code refinement\\
\hline
Sys1 & \tl{Seq2Seq} & \tl{PBSMT} & \tll{LSTM} \\
Sys2 & \tl{Seq2Action+MAML$^1$} & \tl{Transformer} & \tll{Transformer}\\
Sys3 & \tl{GPT2$^2$} & \tl{Transformer+CodeBERT$^4$} & \tll{Transformer+CodeBERT$^4$} \\
Sys4 & \tl{CodeGPT$^3$} & \tl{Human} & - \\

\bottomrule
\end{tabular}}
\end{center}
\caption{\label{tab:exp_systems}
The systems we choose for each task. Note that ``Human" in this table means the output is given by human programming experts. $^1$ \cite{guo2019coupling}; $^2$  Fine-tune with GPT-2 \cite{radford2019language}; $^3$ Pre-trained GPT-2 with the Java data of Codesearchnet \cite{husain2019codesearchnet} and then fine-tuning; $^4$ Fine-tune with CodeBERT \cite{feng2020codebert}.}
\end{table*}

\subsection{Task Introduction}
The three tasks we choose for the experiment are text-to-code, code translation, and code refinement.

\subsubsection{Text-to-code}
Text-to-code \cite{iyer2018mapping} is the task of generating class member functions given the function documentation and the programmatic context. The inputs are the natural language documentation, and the class
environment the code resides in. The environment
comprises two lists of entities: (1) class member
variable names with their data types, and
(2) member function names together with their return types. The output is a piece of code of the desired class member function. We use the same dataset released by \citet{iyer2018mapping}, which consists of 100k training samples, 2k validation samples and 2k test samples.

\subsubsection{Code Translation}
Code translation aims to migrate legacy software from one programming language in a platform to another. Following \citet{nguyen2015divide} and \citet{chen2018tree}, we conduct experiments on a dataset crawled from several open-source projects, i.e., Lucene\footnote{http://lucene.apache.org/}, POI\footnote{http://poi.apache.org/},  JGit\footnote{https://github.com/eclipse/jgit/}, and Antlr\footnote{https://github.com/antlr/}. Those projects have both Java and C\# implementation. We paired the methods in the two languages based on their file names and method names. After removing duplication, the total number of method pairs is 11.8k, and we split 0.5k pairs from them as the development set and another 1k pairs for test. We will release the code translation dataset with our scripts.

\subsubsection{Code Refinement}
Code refinement aims to automatically fix bugs in the code, which can contribute to reducing the cost of bug-fixing for developers. We use the dataset released by \citet{tufano2019empirical}. The source is buggy Java functions while the target is the according fixed ones. Their dataset contains two subsets ( i.e. \textit{small} and \textit{medium}) based on the code length. For the \textit{small} dataset, the function numbers of training, development and test samples are 46,680, 5,835 and 5,835. For the \textit{medium} dataset, the function numbers are 52,364, 6,545 and 6,545 respectively.

\subsection{Settings}
For each task, we prepare 3 to 4 standard systems as shown in Table \ref{tab:exp_systems}. We randomly choose 500 samples from each test set for evaluation. As for human evaluation, we have a group of human judges consisting of 10 people who are familiar with Java and {C\#}. The humans judge our four systems on a subset of 50 samples extracted randomly from our test set. We pair each input with its 4 outputs, resulting in a total of 200 pairs of the given inputs and the output codes. We prepare a UI software with these input-output pairs randomly ordered to disperse the 4 outputs of each input. All judges use this same software and see the pairs in the same order. They rated each output from 1 (very bad) to 5 (very good).


\subsection{Results}
\subsubsection{Main Results}

\begin{table}[ht]
\begin{center}
\setlength{\tabcolsep}{1mm}{
\begin{tabular}{c|c|c|c|c}
\toprule
\multicolumn{5}{c}{Text-to-code} \\
\midrule
System & BLEU & Acc (100\%) & CodeBLEU & Human score \\
\hline
Sys1 & 12.02 & 3.05 & 18.04 & 1.888 \\
Sys2 & 16.82 & 10.50 & 21.71 & 1.99 \\
Sys3 & 21.18 & 17.35 & 24.95 & 2.558 \\
Sys4 & 26.45 & 20.10 & 30.96 & 3.125 \\
\midrule
\multicolumn{5}{c}{Code translation} \\
\midrule
System & BLEU & Acc (100\%) & CodeBLEU & Human score \\
\hline
Sys1 & 44.53 & 13.2 & 45.71 & 3.25 \\
Sys2 & 54.84 & 31.75 & 61.14 & 3.771 \\
Sys3 & 80.18 & 60.2 & 82.74 & 4.036 \\
Sys4 & 81.14 & 63.5 & 84.75 & 4.252 \\
\midrule
\multicolumn{5}{c}{Code refinement} \\
\midrule
System & BLEU & Acc (100\%) & CodeBLEU & Human score \\
\hline
Sys1 & 90.35 & 3.00 & 80.81 & 1.378 \\
Sys2 & 91.40 & 7.01 & 82.16 & 1.545 \\
Sys3 & 92.80 & 17.6 & 83.85 & 2.022 \\
\bottomrule
\end{tabular}}
\end{center}
\caption{\label{tab:exp_main}
The results of all baselines of the given three tasks evaluated by BLEU, accuracy (exactly match), CodeBLEU and human evaluation scores.}
\end{table}

\begin{table*}[ht]
\begin{center}
\begin{tabular}{c|c|c|c|c|c|c|c|c|c}
\toprule
& \multicolumn{3}{c|}{Text-to-code} & \multicolumn{3}{c|}{Code translation} & \multicolumn{3}{c}{Code refinement} \\
\hline
System & Mean & StdDev & t & Mean & StdDev & t & Mean & StdDev & t \\
\hline
Sys1 & 17.93 & 1.8 & - & 44.62 & 5.2 & - & 79.21 & 5.6 & - \\
Sys2 & 20.67 & 2.9 & 7.4 & 60.04 & 5.8 & 30 & 81.04 & 5.8 & 2.1\\
Sys3 & 23.92 & 3.4 & 7 & 81.55 & 6.1 & 38 & 82.52 & 6.4 & 3.4 \\
Sys4 & 30.13 & 4.2 & 12 & 83.26 & 6.7 & 5.2 & - & - & - \\
\bottomrule
\end{tabular}
\end{center}
\caption{\label{tab:exp_var}
The mean, standard deviation and paired t-statistic of all baselines of the given three tasks. The t-statistic compares each system with the neighbor above it in the table. }
\end{table*}

The main results are shown in Table \ref{tab:exp_main}. In this table, we calculate BLEU scores, perfect accuracy, CodeBLEU and human evaluation scores for all systems of each task on the selected test set. Note that the former three metrics are ranging from 0 to 100 and the last one is ranging from 1 (very bad) to 5 (very good). We find that some of the systems are very close in terms of BLEU and CodeBLEU scores. Hence, some questions are raised.

\begin{itemize}
    \item Is the difference in CodeBLEU metric reliable?
    \item What is the variance of the CodeBLEU score?
    \item Is CodeBLEU more correlated with human scores than BLEU and accuracy?
\end{itemize}

To answer these questions, first, following \citet{papineni2002bleu}, we divided the test set into 20 blocks of 25 sentences each, and computed CodeBLEU on these blocks individually. We thus have 20 samples of these metrics for each system. We computed the means, variances, and paired t-statistics for them, which is displayed in Table \ref{tab:exp_var}. 

From Table \ref{tab:exp_var}, as expected, these two sets of results are close for each system and differ only by small finite block size effects. Since a paired t-statistic of 1.7 or above is 95\% significant, the differences between the systems’ scores are statistically very significant. The reported variance on 25-sentence blocks serves as an upper bound to the variance of sizeable test sets like the 500 sentence corpus. Therefore, we conclude that the difference in the CodeBLEU metric is reliable, and the variance of it is within a reasonable range.

\begin{table}[ht]
\begin{center}
\setlength{\tabcolsep}{1mm}{
\begin{tabular}{c|c|c|c}
\toprule
 & Text-to-code & Code trans & Code ref \\
\midrule
BLEU \& human & 0.967 & 0.940 & 0.923 \\
\hline
Acc \& human & 0.912 & 0.968 & \textbf{0.999} \\
\hline
\multirow{2}{*}{CodeBLEU \& human} & \textbf{0.977} & \textbf{0.970} & 0.979 \\
& (+1.0) & (+3.0) & (+5.6) \\
\bottomrule
\end{tabular}}
\end{center}
\caption{\label{tab:exp_pearson}
Comparison of the Pearson correlation coefficients between human evaluation scores and three different metrics. The numbers in the brackets in the last row are the improvements in percent compared with BLEU. }
\end{table}

Next, we compare the correlation of BLEU, accuracy and CodeBLEU to human evaluation scores respectively. The Pearson correlation coefficients are listed in Table \ref{tab:exp_pearson}. 

From the table, we see CodeBLEU scores are more correlated with human evaluation scores in all the three tasks. The improvements are significant compared with the traditional MT metric BLEU. The results verify the effectiveness of our proposed metric. For text-to-code and code translation tasks, CodeBLEU scores are also more correlated with human scores than accuracy (Acc), but there is an exception that the Acc is more correlated for code refinement. This is because the data of refinement task is just fixing small bugs in a given Java function. The output is usually unique, and the humans score the outputs based on the unique refinement way, so that the Acc here correlates more with human evaluation scores. However, we also believe that in the more general code synthesis scenarios, CodeBLEU is more reasonable in terms of the correlation with human scores.

\begin{figure}[!h]
	\centering
	\includegraphics[width=8cm]{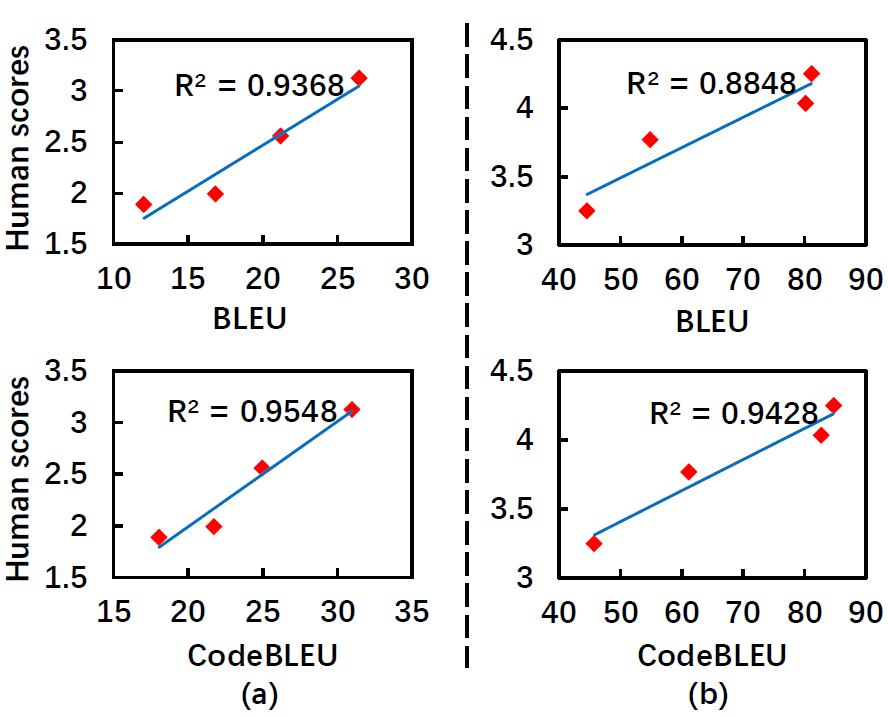}
	\caption{BLEU and CodeBLEU predict human evaluation scores. (a) Text-to-code; (b) Code translation.
	}\label{fig:exp_regression}
\end{figure}

Figure \ref{fig:exp_regression} shows the comparable regression results for each metric to human scores on the text-to-code and code translation tasks. The $R^2$ values of the linear regression are also shown in the figure. From the figure, we find CodeBLEU is more linearly correlated with human evaluation scores than BLEU, which is consistent with the results in Table \ref{tab:exp_pearson}. 

Based on the above results and analysis, we conclude that:
\begin{itemize}
    \item The difference in CodeBLEU metric is reliable. CodeBLEU is capable to differentiate code synthesis systems.
    \item CodeBLEU is reliable, and its variance is within a reasonable range. 
    \item CodeBLEU is more correlated with human evaluation scores than traditional BLEU scores on all the three tasks, and more correlated than Acc on the two tasks.
\end{itemize}

\subsubsection{Ablation Study}
To investigate the influence of the different components of CodeBLEU, we conduct the following experiment to calculate the respective Pearson correlation between  the human evaluation scores and the scores given by different components. The results are reported in Table \ref{tab:exp_ablation}.

\begin{table}[ht]
\begin{center}
\setlength{\tabcolsep}{1mm}{
\begin{tabular}{c|c|c|c}
\toprule
Components & Text-to-code & Code trans & Code ref \\
\hline
$\rm{ BLEU}$ & 0.967 & 0.940 & 0.923  \\
$\rm{ BLEU_{weight}}$ & 0.960 & 0.934 & 0.985 \\
$\rm{ Match_{ast}}$ & 0.985 & 0.977 & 0.967 \\
$\rm{ Match_{df}}$ & 0.978 & 0.974 & 0.983 \\
\hline
$\rm{CodeBLEU}$ & 0.977 & 0.970 & 0.979 \\
\bottomrule
\end{tabular}
}
\end{center}
\caption{\label{tab:exp_ablation}
The Pearson correlation coefficients between different components of CodeBLEU and humans.}
\end{table}


From the table, we find that, for the text-to-code and code translation tasks, the scores of the last two components, i.e., syntactic AST match and semantic data-flow match, are more relevant to human evaluation scores compared with the n-gram and weight n-gram match scores. For the code refinement task, the scores given by the weighted n-gram match and the semantic data-flow are more relevant to human evaluation. This may be because many bugs in the refinement training data are wrong variable naming or keywords errors, while the weighted n-gram and semantic data-flow match scores could evaluate them better. The above result verifies the effectiveness of our three proposed components, i.e., weighted n-gram match, syntactic AST match and semantic data-flow match, for code synthesis evaluation. Besides, the results are inspiring for us to change the hyper-parameters $\alpha,\beta,\gamma,\delta$ in Eq. (\ref{eq:CodeBLEU}) to get better evaluation whose results are more correlated with humans. For example, to achieve this, we can increase $\gamma$ and $\delta$ to improve the weights of the last two components in the final CodeBLEU scores. In the next section, we will conduct experiments to investigate the influence of the four hyper-parameters.

\subsection{Influence of hyper-parameters}
\begin{figure}[!h]
	\centering
	\includegraphics[width=7cm]{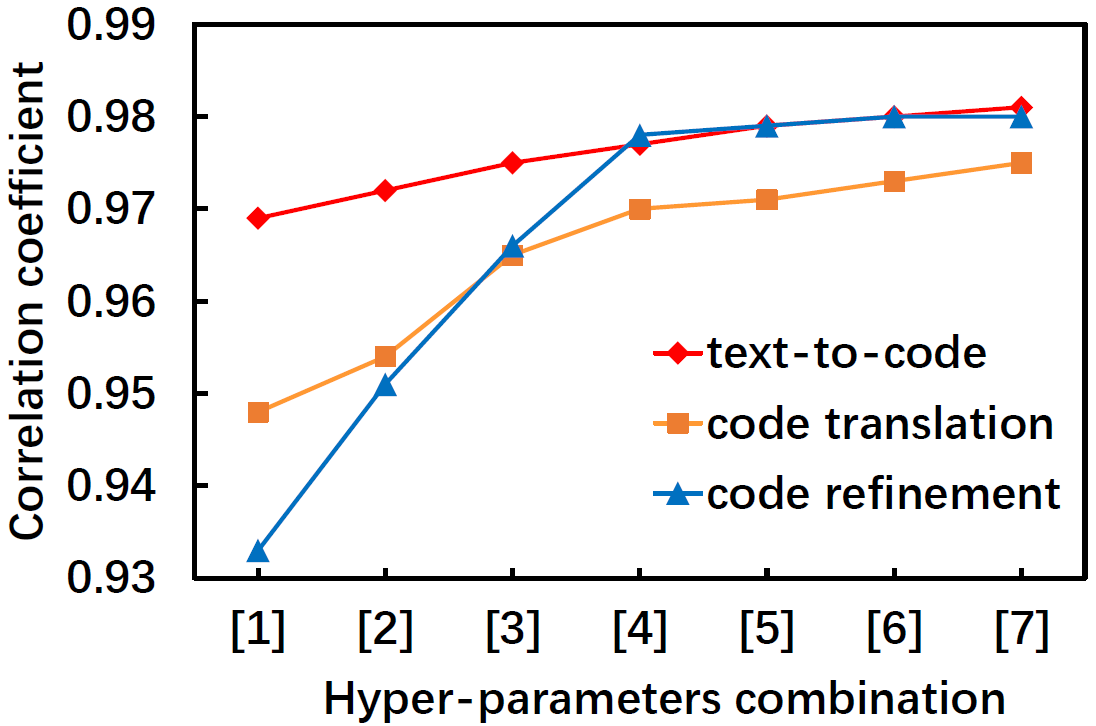}
	\caption{The correlation coefficients between CodeBLEU and human scores with different hyper-parameters. The hyper-parameter setting of each combination is in Table \ref{tab:exp_params_setting}.
	}\label{fig:exp_hyp-params}
\end{figure}

In the above subsection, we find different components have a different influence on the final results of CodeBLEU in terms of the correlation with human evaluation scores. Therefore, we can change the weights of those components to achieve a higher correlation between CodeBLEU and human evaluation. We gradually increase the weights of the last two components (as in Table \ref{tab:exp_params_setting}) and record the correlation coefficients between CodeBLEU and human evaluation scores for the three tasks. The results are shown in Figure \ref{fig:exp_hyp-params}. 

From the figure, we find that increasing the weights of the last two components improves the correlation between CodeBLEU and human scores for all of the three tasks. The performance starts to converge after the combination [4] and the combination [7], i.e., $\alpha, \beta, \gamma, \delta = 0.1, 0.1, 0.4, 0.4$, achieves the best result among all the combinations in Figure \ref{fig:exp_hyp-params} (0.981, 0.975, 0.980 for the three tasks respectively). Of course, [7] is not the best combination all the time. For example, $\alpha, \beta, \gamma, \delta = 0.1, 0.4, 0.1, 0.4$ achieves the better result (the correlation coefficient is 0.984) than the combination [7] (the correlation coefficient is 0.980) for the code refinement task. In spite of this, we recommend to choose the combination [7] when calculating CodeBLEU for general code synthesis tasks, because the last two components are more likely to be more correlated with human evaluation scores from the instinct given by Table \ref{tab:exp_pearson}.

\begin{table}[ht]
\begin{center}
\begin{tabular}{c|c}
\hline
Combination & $\alpha, \beta, \gamma, \delta$\\
\hline
[1] & 0.40, 0.40, 0.10, 0.10 \\
\hline
[2] & 0.35, 0.35, 0.15, 0.15 \\
\hline
[3] & 0.30, 0.30, 0.20, 0.20 \\
\hline
[4] & 0.25, 0.25, 0.25, 0.25 \\
\hline
[5] & 0.20, 0.20, 0.30, 0.30 \\
\hline
[6] & 0.15, 0.15, 0.35, 0.35 \\
\hline
[7] & 0.10, 0.10, 0.40, 0.40 \\
\hline
\end{tabular}
\end{center}
\caption{\label{tab:exp_params_setting}
The settings of each combination in Figure \ref{fig:exp_hyp-params}.}
\end{table}

\section{Related Work}
As code artificial intelligence receives more and more attention ~\cite{allamanis2015bimodal,yin-neubig-2017-syntactic,allamanis2018survey,monperrus2018automatic,alon2019code2vec,svyatkovskiy2020intellicode}, the evaluation of code synthesis becomes critical to promote its development.
Although there are several automatic evaluation methods, which can be used to evaluate code synthesis \cite{karaivanov2014phrase,chen2018tree,lachaux2020unsupervised}, these approaches still suffer from many weakness and are not suitable to evaluate code. 

The widely used 4-gram BLEU \cite{papineni2002bleu} evaluates the code quality by using the relative overlap between the tokens in the hypothesis and reference \cite{karaivanov2014phrase,barone2017parallel}.
Nevertheless, BLEU ignores the grammatical correctness and logic correctness.
The perfect accuracy \cite{rabinovich2017abstract,chen2018tree} is too strict and it is an underestimation of the true accuracy based on semantic equivalence.
Additionally, the computational accuracy \cite{lachaux2020unsupervised}, evaluating whether the hypothesis function generates the same outputs given the same inputs by performing code, locks universality and practicability.
To overcome the limitation, our proposed simple and effective CodeBLEU can not only consider the surface match similar with the original BLEU, but can also consider the grammatical correctness and the logic correctness.

\section{Conclusion}
In this paper, we propose a novel metric CodeBLEU for code synthesis evaluation. CodeBLEU evaluates the candidate code pieces considering not only the shallow match, but also the syntactic match and the semantic match. The results of three real-world tasks, i.e. text-to-code, code translation and code refinement, demonstrate the rationality and effectiveness of CodeBLEU by analyzing the correlation with human evaluation scores from different granularity. In the future work, we will delve more into the evaluation of syntactic and semantic match and try more tasks with CodeBLEU to show its practicality.

\bibliography{aaai}
\bibliographystyle{aaai}

\end{CJK*}
\end{document}